\documentclass[pra,
superscriptaddress,
amssymb,amsmath,amsmath,showpacs,reprint,twocolumn]{revtex4-1}
\usepackage{color}
\usepackage{graphicx}
\usepackage{epstopdf}
\usepackage{natbib}
\usepackage{wasysym}
\usepackage{hyperref}
\usepackage{dcolumn}

\hypersetup{%
   pdfpagemode=None, %FullScreen,
   pdfstartpage=1,
   pdfstartview=FitH,
   pdfmenubar=true,
   pdftoolbar=true,
   colorlinks = true,
   linkcolor=blue,
   citecolor=blue,
   bookmarksopen=false
}

\def\half{\frac{1}{2}}

\usepackage{booktabs}
\usepackage{multirow}
\usepackage{graphicx}
\usepackage{amsmath}
\usepackage{indentfirst}
\begin{document}
\title{Potential energy curve for the $a^3\Sigma_u^+$ state of lithium dimer with Slater-type orbitals}

\author{\sc Micha\l\ Lesiuk}
\email{e-mail: lesiuk@tiger.chem.uw.edu.pl}
\affiliation{\sl Faculty of Chemistry, University of Warsaw, Pasteura 1, 02-093 Warsaw, Poland}
\author{\sc Monika Musia\l}
\affiliation{\sl Institute of Chemistry, University of Silesia in Katowice, Szkolna 9, 40-006 Katowice, Poland}
\author{\sc Robert Moszynski}
\affiliation{\sl Faculty of Chemistry, University of Warsaw, Pasteura 1, 02-093 Warsaw, Poland}
\date{\today}

\begin{abstract}
We report state-of-the-art \emph{ab initio} calculations of the potential energy curve for the $a^3\Sigma_u^+$ state of
the lithium dimer conducted to achieve spectroscopic accuracy ($<$1cm$^{-1}$) without any prior adjustment to fit the
corresponding experimental data. The nonrelativistic clamped-nuclei component of the interaction energy is
calculated with a composite method involving six-electron coupled cluster and full configuration interaction theories
combined with basis sets of Slater-type orbitals ranging in quality from double- to sextuple-zeta. 
We additionally include both the leading-order relativistic and adiabatic corrections, and
find both of these effects to be non-negligible within the present accuracy standards.
The potential energy curve developed by us allowed to calculate molecular parameters ($D_e$, $D_0$, 
$\omega_e$ etc.) for this system, as well as the corresponding vibrational energy levels, with an 
error of only about $0.2-0.4\,$cm$^{-1}$. We also report an \emph{ab initio} value for the 
scattering length of two $^2S$ lithium atoms.
\end{abstract}

\maketitle

\section{Introduction}
\label{sec:intro}

Lithium dimer is one of the simplest (bound) homonuclear many-electron molecules. Therefore, it has attracted 
a significant attention in the past years with many experimental
\cite{szmek77,kusch77,hessel79,bernheim81a,bernheim81b,bernheim82,bernheim83,engelke83,xie85,xie86,bernheim87,linton89a,
linton89b,ishikawa91,miller90,che91,lyyra91,linton93,yianna94,yianna95,antonova96,ross98,martin98,linton99,antonova00,
kasahara00,lazarov01,bauloufa01,song02,kubkowska07,lazarov08,gunton13,semczuk13} and theoretical
\cite{cooper82,partridge83,bishop84,chung99,pashov00,jastrzebski01,shi07} works devoted 
entirely to its observation and description. However, singlet electronic states of Li$_2$ were the 
main subjects of the studies; Refs. \cite{barakat86,linton96,martin97,wang98,urbanski02,wang02,adohi04,coxon06,leroy09}
provide a good overview on this topic. 

In contrast, the triplet electronic states of the lithium dimer have been observed for the first 
time only relatively recently. Experimental studies of the triplet states of Li$_2$ are difficult
because transitions from the ground $X^1\Sigma_g^+$ state are dipole-forbidden. Moreover, the spin-orbit coupling in
lithium
is very weak. This impasse has been broken by improvements in experimental techniques such as perturbation-facilitated
optical-optical double resonance (PFOODR) \cite{yianna94,yianna95,xie85,xie86}. Xie and Field \cite{xie85,xie86} were
the first to access the triplet
state $a^3\Sigma_u^+$ and determine the relevant spectroscopic constants. They started with the (bound) ground state
$X^1\Sigma_g^+$ and excited into the mixed $A^1\Sigma_u^+ - b^3\Pi_u$ manifold. A subsequent fluorescence led to the
final $a^3\Sigma_u^+$ state. Later, Martin et al. \cite{martin97,martin98}, Linton et al.
\cite{linton89a,linton89b}, and others \cite{lyyra91} determined accurate vibrational and rotational constants
for this state by using high-resolution Fourier transform spectroscopy. These data were
further revised by Zemke and Stwalley \cite{zemke93} reporting more bound vibrational levels than initially claimed.
Abraham et al. \cite{abraham95} performed photoassociation of ultracold lithium atoms allowing to determine
precise positions of the highest vibrational levels. Finally, Linton et al. \cite{linton99} determined
spectroscopic constants for the $a^3\Sigma_u^+$ state to an accuracy of only a small fraction of 
cm$^{-1}$. This
progress was accompanied by a number of works where semiempirical potentials were developed to reproduce the
experimental spectra (see, for example, Refs. \cite{moerdijk94,cote94,dattani11,lau16} and references therein).

Observation of the Bose-Einstein condensate of the lithium atoms \cite{abraham95,jochim03,chin04} sparked a renewed
interest in the $a^3\Sigma_u^+$ electronic state, also in analogous diatomic molecules composed of 
heavier alkali metals \cite{gronowski20}. The reason is the relation between the stability of the 
Bose-Einstein condensate of spin-polarized atoms and the scattering length ($a$) of these atoms. 
This 
quantity can be calculated from first principles having an accurate potential energy curve (PEC) for 
the $a^3\Sigma_u^+$ state. Unfortunately, the scattering length is very sensitive to tiny details of 
the PEC, especially in the asymptotic region. This can be illustrated by an approximate formula 
\cite{roman65}, $a^2 \approx \frac{\hbar^2}{m |E_b|}$, relating the scattering length $a$ to the 
binding energy of the highest occupied vibrational level, $E_b$ ($m$ is the atomic mass).
One can see that even a relatively small change in the well-depth of the PEC can shift the value
of $E_b$ significantly and thus impact the calculated scattering length dramatically. This makes 
accurate \emph{ab initio} determination of $a$ very challenging and it has been achieved thus far 
only for the smallest systems. Quite recently, lithium atom and dimer has been also the subject of 
research in the context of quantum information 
theory~\cite{benavides13,theophilou15,benavides15,reiher18}.

The triplet $a^3\Sigma_u^+$ state of the lithium dimer is weakly bound with the PEC well-depth of about 
334 cm$^{-1}$ and a minimum around $4.2\,$\AA{} \cite{dattani11}. Despite that, it accommodates as many as ten
vibrational 
levels. To get a broader picture, let us present a short survey of theoretical results available in the 
literature for this state. 

The first works devoted to various electronic states of Li$_2$ employed effective core potentials (with one valence
electron) and optional core polarization corrections. The papers of Konowalow and coworkers
\cite{olson76,olson77a,olson77b,konowalow79,konowalow83}, M\"{u}ller and Mayer \cite{muller84},
Schmidt-Mink et al. \cite{schmidt85}, and several others \cite{poteau95}, are prime examples of this approach.
The biggest advantage of the core potentials is that the remaining effective two-electron system can be
treated with relative ease.
As a result, many excited states of different spatial and spin symmetries can be studied simultaneously, as best
illustrated by recent papers of Jasik et al. \cite{jasik06,jasik07,jasik13}. Unfortunately, the accuracy of
this effective approach is somewhat limited, with errors reaching several percents for some quantities. To reduce this
error a more elaborate first-principles method must be used. This has recently been achieved by Musial and Kucharski
\cite{musial14} by using a sophisticated all-electron coupled cluster approach. The error has been reduced by an order
of magnitude compared with the previous works; at the same time, more than thirty electronic states 
were characterized.

In this paper we present state-of-the-art \emph{ab initio} PEC for the
$a^3\Sigma_u^+$ state of the lithium dimer. We combine high-level quantum chemical methods with 
large one-electron basis sets composed of Slater-type orbitals (STOs) \cite{slater30,slater32} to reach saturation of
the calculated values. We employ techniques for calculation of the two-center matrix elements over 
STOs reported
recently \cite{lesiuk12,lesiuk14a,lesiuk14b,lesiuk15,lesiuk16}. Moreover, we evaluate corrections arising from several
minor physical effects, e.g., adiabatic or
relativistic. We also calculate various spectroscopic parameters such as dissociation energy, vibrational energy
levels etc. and compare them with the latest experimental data. We would like to emphasize that all 
calculations reported here utilize only rigorous \emph{ab initio} methods. In other words, the 
results were obtained
with no prior reference to the empirical data.

Atomic units are used throughout the paper unless explicitly stated otherwise. We adopt the
following conversion factors and fundamental constants: $1\,a_0 = 0.529\,177$ \AA{} (Bohr radius), 
$1\,\mbox{u}=1822.888$ (unified atomic mass unit), $1\,$H=$219\,474.63$ cm$^{-1}$ (Hartree),
$\alpha$ = $1 / 137.035\,999$ (the fine structure constant). These values are in line with the 
recent CODATA recommendations \cite{mohr16}. We also adopt a convention that the interaction energy 
is positive whenever the underlying interaction is attractive.

\section{Electronic structure calculations}
\label{sec:abinit}

\subsection{Basis sets}
\label{subsec:basis}

In accurate \emph{ab initio} calculations employing basis sets of any kind it is of uttermost 
importance to generate a systematic sequence of basis sets guaranteeing that the results 
converge to the exact answer. This allows for reliable extrapolation towards the complete basis 
set (CBS) limit and (partly) overcomes the slow convergence of the correlation energy with the basis set size. 
Unfortunately, we are not aware of any openly available Slater-type basis sets which would satisfy 
the present accuracy requirements. There are many papers devoted to optimization of the STOs basis 
sets in the literature \cite{clementi74,mclean81,snijders81,lenthe03,chong04,chong05}. However, they are either very old
and concentrated mainly on atomic 
properties or aimed at the density functional theory calculations where the basis set requirements 
are different. As a result, the first step of this work is optimization of Slater-type basis sets 
fulfilling the high accuracy standards of the present study.

All basis sets used in this paper are composed of canonical STOs \cite{slater30,slater32}
\begin{align}
\label{sto1}
\chi_{lm}(\textbf{r};\zeta) = \frac{(2\zeta)^{n+1/2}}{\sqrt{(2n)!}}\, r^l e^{-\zeta r}\,
Y_{lm}(\theta,\phi),
\end{align}
where $\zeta>0$ is a free nonlinear parameter, and $Y_{lm}$ are the spherical harmonics in 
the Condon-Shortley phase convention. By the term ``canonical STOs'' we mean that the power of $r$ 
is equal to the angular momentum, $l$.

\begin{table}[t]
\caption{Composition of the STOs basis sets wtcc-$l$ and da-wtcc-$l$ for the lithium atom; 
$l$ is the largest angular momentum included (see the main text for details).}
\begin{ruledtabular}
\begin{tabular}{c|ll}
\label{bas}
$l$ & atomic & diffuse \\
\hline\\[-2.1ex]
1 & $5s1p$ & $2s1p$ \\[0.6ex]
2 & $6s2p1d$ & $2s2p1d$ \\[0.6ex]
3 & $7s3p2d1f$ & $2s2p2d1f$ \\[0.6ex]
4 & $8s4p3d2f1g$ & $2s2p2d2f1g$ \\[0.6ex]
5 & $9s5p4d3f2g1h$ & $2s2p2d2f2g1h$ \\[0.6ex]
6 & $10s6p5d4f3g2h1i$ & $2s2p2d2f2g2h1i$ \\[0.6ex]
\end{tabular}
\end{ruledtabular}
\end{table}

To optimize the nonlinear parameters we employ the well-tempering 
scheme~\cite{huzinaga85a,huzinaga85b,huzinaga90}; exponents for a given 
angular momentum $l$ are written as
\begin{align}
\label{wt}
\zeta_{lk} = \alpha_l\, \beta_l^{k + \gamma_l k^2} \;\;\; \mbox{with}\; k=0,1,2,\ldots
\end{align}
where $\alpha_l$, $\beta_l$, and $\gamma_l$ are the actual parameters which have to be determined 
variationally. Well-tempering (or related schemes) not only reduce the computational costs of the 
optimization, but also alleviate the linear dependency problems and help to avoid troublesome 
local minima. The latter merit is particularly advantageous in maintaining the consistency of 
the basis sets sequence. At the same time, the flexibility of Eq. (\ref{wt}) is usually surprisingly good.
Brute-force optimizations typically give only marginally better results, especially when a large 
number of functions are included.

When deciding on the composition of the STOs basis sets we follow the correlation-consistency 
principle, first proposed by Dunning \cite{dunning89}.
The smallest basis set considered here has 
the composition $5s1p$ and is systematically expanded, re-optimizing the nonlinear parameters at 
each step. This gives a sequence of basis sets denoted shortly wtcc-$l$ (well-tempered 
correlation-consistent) where $l$ is the largest angular momentum included. A 
detailed composition of these basis sets is given in Table \ref{bas}. To find the optimal values of 
the well-tempering parameters for each $l$ we minimized the total 
CISD energy of the lithium atom with all electrons active.

Basis sets designed to reproduce the atomic energies may not 
be equally satisfactory in a molecular environment. This is especially true for weakly bound
systems where the tails of the electronic density are important for the bonding phenomena. To assure 
that the basis sets developed here are 
truly universal we supplemented them with two sets of diffuse functions, see Table \ref{bas}. The 
exponents of these functions were varied freely to maximize the static dipole polarizability of the 
lithium atom evaluated at the coupled Hartree-Fock level of theory. The modified (augmented) basis 
sets are denoted da-wtcc-$l$ where ``da'' stands for doubly augmented. 

Finally, in this work we are concerned with the calculation of the relativistic corrections 
which have somewhat specific basis set requirements. To eliminate possible sources of error we created 
a special sequence of basis sets denoted (da-)wtcc-$l$+s. These basis sets share the polarization 
and/or augmented functions with the standard (da-)wtcc-$l$, but all $s$ functions were replaced 
with a universal set of twelve 1$s$ orbitals obtained by minimizing the Hartree-Fock energy of the 
lithium atom. Detailed compositions of all basis sets used in this work (including values of the nonlinear parameters)
are given in Supplemental Material \cite{supplement}.

\begin{table}[t]
\caption{Total energy ($E_{\mbox{\scriptsize total}}$) and the correlation energy 
($E_{\mbox{\scriptsize corr}}$) of the lithium atom calculated at the FCI level of theory by using 
the STOs basis sets da-wtcc-$l$. All values are given in the atomic units.}
\begin{ruledtabular}
\label{tatom}
\begin{tabular}{cccc}
$l$ & $E_{\mbox{\scriptsize corr}}$ & $E_{\mbox{\scriptsize total}}$ \\
\hline\\[-2.1ex]
2 & $-$0.041 842 & $-$7.474 511 \\[0.6ex]
3 & $-$0.043 749 & $-$7.476 454 \\[0.6ex]
4 & $-$0.044 532 & $-$7.477 239 \\[0.6ex]
5 & $-$0.044 862 & $-$7.477 569 \\[0.6ex]
6 & $-$0.045 056 & $-$7.477 763 \\[0.6ex]
\hline\\[-2.1ex]
$\infty$ & $-$0.045 386 & $-$7.478 093 \\[0.6ex]
\hline\\[-2.1ex]
Ref. \cite{puchalski06}  & $-$0.045 353 & $-$7.478 060 \\
\end{tabular}
\end{ruledtabular}
\end{table}

As a benchmark of the newly developed basis sets we compared our atomic results with the reference 
values available in the literature. For the lithium atom very accurate value of the 
clamped-nucleus nonrelativistic energy is available \cite{puchalski06} from the three-body Hylleraas calculations, 
$E_{\mbox{\scriptsize total}}=-7.478 060 323 904 1{+10\choose -50}$. This value is 
virtually exact for the present purposes. For comparison, we calculated Hartree-Fock and FCI 
correlation energies in the da-wtcc-$l$ basis sets, see Table \ref{tatom}.

The Hartree-Fock energy converges at an exponential rate. Indeed, by comparing the results from 
the largest two basis sets we see that the energy difference is less than 1$\mu$H. Therefore, we 
simply take the value from the largest basis set, $E_{\mbox{\scriptsize 
HF}}=-$7.432 707(1), and conservatively assume that the error is at most 1$\mu$H. Extrapolation of the
HF energies by using the exponential formula barely changed the results. On the other 
hand, the correlation energy converges at a much slower rate and we apply the 
conventional three-point extrapolation~\cite{hill85}
\begin{align}
\label{extra}
E = a + \frac{b}{l^3} + \frac{c}{l^5},
\end{align}
where the constants $a,b,c$ are obtained by fitting. In Table \ref{tatom} we present results 
obtained with the basis sets $l=2-6$ and the values obtained by the extrapolation.
Note that our final number for the total energy of the lithium atom differs by only about 34 $\mu$H 
($\approx$7 cm$^{-1}$) from the aforementioned reference value.

\subsection{Born-Oppenheimer potential}
\label{subsec:bopot}

\begin{table}[t]
\caption{Nonrelativistic contributions to the interaction energy of the lithium dimer calculated 
with the da-wtcc-$l$ basis sets; $E_{\mbox{\scriptsize int}}^{\mbox{\scriptsize HF}}$ and 
$E_{\mbox{\scriptsize int}}^{\mbox{\scriptsize ccsd(t)}}$ denotes the interaction energy obtained 
at Hartree-Fock and CCSD(T) level of theory, respectively. The abbreviations $\Delta 
E_{\mbox{\scriptsize int}}^{\mbox{\scriptsize ccsdt}}$ and $\Delta E_{\mbox{\scriptsize 
int}}^{\mbox{\scriptsize fci}}$ stand for the post-CCSD(T) corrections, see Eqs. (\ref{corrt}) and 
(\ref{corrfci}). Counterpoise correction was applied to remove the basis set superposition error. 
All values are given in cm$^{-1}$.}
\begin{ruledtabular}
\label{tdim}
\begin{tabular}{ccccc}
$l$       & $E_{\mbox{\scriptsize int}}^{\mbox{\scriptsize HF}}$ 
          & $E_{\mbox{\scriptsize int}}^{\mbox{\scriptsize ccsd(t)}}$
          & $\Delta E_{\mbox{\scriptsize int}}^{\mbox{\scriptsize ccsdt}}$
          & $\Delta E_{\mbox{\scriptsize int}}^{\mbox{\scriptsize fci}}$ \\[0.6ex]
\hline\\[-2.1ex]
 \multicolumn{5}{c}{$R=7.75$} \\
\hline\\[-2.1ex]
2 & $-$359.46 & 276.40 & 1.45 & 0.15 \\[0.6ex]
3 & $-$345.15 & 322.50 & 2.28 & ---  \\[0.6ex]
4 & $-$344.14 & 328.30 & 2.29 & ---  \\[0.6ex]
5 & $-$344.05 & 329.21 & ---  & ---  \\[0.6ex]
6 & $-$344.10 & 329.51 & ---  & ---  \\[0.6ex]
\hline\\[-2.1ex]
$\infty$      & $-$344.05$\,\pm\,$0.01 & 330.09$\,\pm\,$0.29 & 2.30$\,\pm\,$0.12 & 
0.18$\,\pm\,$0.05 \\
\hline\\[-2.1ex]
\multicolumn{5}{c}{$R=12.5$} \\
\hline\\[-2.1ex]
2 & $-$9.64 & 73.79 & 0.22 & 0.05 \\[0.6ex]
3 & $-$9.51 & 85.83 & 0.30 & --- \\[0.6ex]
4 & $-$9.46 & 87.05 & 0.30 & --- \\[0.6ex]
5 & $-$9.45 & 87.26 & ---  & --- \\[0.6ex]
6 & $-$9.44 & 87.74 & ---  & --- \\[0.6ex]
\hline\\[-2.1ex]
$\infty$      & $-$9.42$\,\pm\,$0.01 & 87.96$\,\pm\,$0.11 & 0.30$\,\pm\,$0.02 & 0.06$\,\pm\,$0.02 \\
\end{tabular}
\end{ruledtabular}
\end{table}

Lithium dimer is a two-center six-electron molecule. For such system the FCI method, which gives 
the exact solution of the Schr\"{o}dinger equation in the CBS limit, cannot be applied. Therefore, 
in the present work we rely on a composite method which is based mostly on the coupled cluster (CC) 
theory~\cite{cizek66,bartlett07}. 

Within the Born-Oppenheimer (BO) approximation, the interaction energy of the lithium dimer for 
each internuclear distance is defined as
\begin{align}
 -E_{\mbox{\scriptsize int}}^{\mbox{\scriptsize X}} = 
 E_{\mbox{\scriptsize X}}(\mbox{Li}_2) - 2\,E_{\mbox{\scriptsize X}}(\mbox{Li})
\end{align}
where $E_{\mbox{\scriptsize X}}(\mbox{Li}_2)$ is the energy of the molecule in the $a^3\Sigma_u^+$ 
state, $E_{\mbox{\scriptsize X}}(\mbox{Li})$ is the ground-state energy of the atom, and the 
superscript X denotes the level of theory. The negative sign in front of the above formula is a 
convention. Unless explicitly stated otherwise, the counterpoise correction~\cite{boys70} is used 
in computation of the interaction energies in order to eliminate the basis set superposition error. 
In this approach the energy of the atom is calculated in the basis set of the molecule and thus the 
quantity $E_{\mbox{\scriptsize X}}(\mbox{Li})$ is 
different for each internuclear separation.  Our protocol for obtaining accurate Born-Oppenheimer 
interaction energies is as follows.

First, we evaluate the BO interaction energies by using the Hartree-Fock 
and CCSD(T) \cite{raghavachari89} methods (all 
electrons active). The values obtained are abbreviated shortly $E_{\mbox{\scriptsize 
int}}^{\mbox{\scriptsize HF}}$ 
and $E_{\mbox{\scriptsize int}}^{\mbox{\scriptsize ccsd(t)}}$, respectively. At these levels of 
theory the complete sequence of basis sets, $l=2-6$, can be used. The Hartee-Fock and 
correlation 
contributions are extrapolated separately - the exponential formula is used for the HF component 
and the formula (\ref{extra}) is applied for the remainder. In Table \ref{tdim} we present 
results of this procedure for two interatomic distances - $7.75$ a.u. and $12.5$ a.u. The former 
value is near the minimum of PEC whilst the latter lies close to the 
dissociation limit.

Interestingly, there is a small inconsistency in the Hartee-Fock values - the interaction energy 
calculated with the $l=6$ basis set is by a tiny bit smaller than with $l=5$. To overcome this 
problem 
we extrapolate the HF limit from the $l=3,4,5$ basis sets, omitting the $l=6$ value. Due to 
comparatively fast convergence of the HF energies towards the CBS limit the error 
introduced by this approximation is minor for all interelectronic distances. More importantly, this 
artifact is absent in the correlated contribution and thus not of a major concern. In the
estimation of the extrapolation errors we adopt a fairly conservative approach. Unless explicitly 
stated otherwise, we assume that the uncertainty is equal to a half of the difference between the 
extrapolated result and the corresponding value in the largest basis set.

To bring the accuracy down to the sub-cm$^{-1}$ regime we need to consider some minor 
corrections beyond the CCSD(T) model. They naturally split 
into two contributions. The first is the full triples correction, being defined as a difference between 
the interaction energies obtained with the CCSDT~\cite{noga87} and CCSD(T) methods
\begin{align}
\label{corrt}
 \Delta E_{\mbox{\scriptsize 
int}}^{\mbox{\scriptsize ccsdt}}=E_{\mbox{\scriptsize int}}^{\mbox{\scriptsize 
ccsdt}}-E_{\mbox{\scriptsize int}}^{\mbox{\scriptsize ccsd(t)}}
\end{align}
The second correction accounts 
for excitations higher than triple and is calculated as a difference between the FCI and CCSDT 
interaction energies
\begin{align}
\label{corrfci}
 \Delta E_{\mbox{\scriptsize int}}^{\mbox{\scriptsize 
fci}}=E_{\mbox{\scriptsize int}}^{\mbox{\scriptsize fci}}-E_{\mbox{\scriptsize 
int}}^{\mbox{\scriptsize ccsdt}}
\end{align}
The post-CCSD(T) corrections are especially computationally intensive. 
In fact, we were able to calculate $\Delta E_{\mbox{\scriptsize int}}^{\mbox{\scriptsize ccsdt}}$ 
in basis sets only up to $l=4$. Even more disappointingly, the FCI correction is feasible only in 
the smallest basis set considered here, $l=2$. These restrictions eliminate the possibility of a 
reliable extrapolation.

To estimate the CBS limits of the post-CCSD(T) corrections we invoke a different strategy. Let us 
assume that the rate of convergence of the interaction energy with respect to the basis set size is 
the same at the CCSD(T) level and for the post-CCSD(T) corrections. Because a reliable limit of 
the CCSD(T) interaction energy is known, approximate CBS limits of the $\Delta 
E_{\mbox{\scriptsize int}}^{\mbox{\scriptsize ccsdt}}$ and $\Delta E_{\mbox{\scriptsize 
int}}^{\mbox{\scriptsize fci}}$ corrections can now be obtained by a simple scaling. The scaling 
parameter is chosen so that the interaction energy calculated with a given 
finite basis set at the CCSD(T) level matches the extrapolated value.

Clearly, the scaling procedure is not as reliable as extrapolation, the latter having firm 
theoretical underpinnings. We assume that this procedure gives an accuracy of 5\% for 
$\Delta E_{\mbox{\scriptsize int}}^{\mbox{\scriptsize ccsdt}}$ and 25\% for 
$\Delta E_{\mbox{\scriptsize int}}^{\mbox{\scriptsize fci}}$.  The results of 
the scaling are 
given in Table \ref{tdim}. The final theoretical error is computed by summing squares of the 
uncertainties in the individual components and taking the square root. For example, at the 
internuclear distance $R=7.75$ this gives 332.48$\,\pm\,$0.28 cm$^{-1}$ for the total 
BO interaction energy.

\subsection{Relativistic effects}
\label{subsec:relpot}

\begin{table}[t]
\caption{Relativistic corrections to the interaction of the lithium dimer energy calculated
with the da-wtcc-$l$ basis sets. The corrections $\langle P_4 \rangle$ and $\langle D_1 \rangle$ 
are defined by Eqs. (\ref{p4}) and (\ref{d1}), respectively. The last column provides
the sum of the values from the preceding two. All values are given in cm$^{-1}$.}
\begin{ruledtabular}
\label{trel}
\begin{tabular}{cccc}
$l$ & $\langle P_4 \rangle$ 
          & $\langle D_1 \rangle$
          & total Cowan-Griffin \\[0.4ex]
%           & $\langle B \rangle$ \\[0.6ex]
\hline\\[-2.1ex]
 \multicolumn{4}{c}{$R=7.75$} \\
\hline\\[-2.1ex]
2 & $-$0.85 & 0.63 & $-$0.22 \\[0.6ex]
3 & $-$0.91 & 0.67 & $-$0.24 \\[0.6ex]
4 & $-$0.91 & 0.67 & $-$0.24 \\[0.6ex]
5 & $-$0.91 & 0.67 & $-$0.24 \\[0.6ex]
\hline\\[-2.1ex]
\multicolumn{4}{c}{$R=12.5$} \\
\hline\\[-2.1ex]
2 & $-$0.11 & 0.08 & $-$0.03 \\[0.6ex]
3 & $-$0.13 & 0.09 & $-$0.04 \\[0.6ex]
4 & $-$0.13 & 0.09 & $-$0.04 \\[0.6ex]
5 & $-$0.14 & 0.10 & $-$0.04 \\[0.6ex]
\end{tabular}
\end{ruledtabular}
\end{table}

For light systems, such as the lithium dimer, the leading-order relativistic corrections (quadratic 
in the fine structure constant, $\alpha$) can be calculated perturbatively. Here we adopt the 
approach based on the one-electron part of the Breit-Pauli Hamiltonian \cite{bethe75}
\begin{align}
\label{breit}
E^{(2)} &= \langle P_4 \rangle + \langle D_1 \rangle,
%  + \langle D_2 \rangle + \langle B \rangle,
\end{align}
\begin{align}
\label{p4}
\langle P_4 \rangle &= -\frac{\alpha^2}{8} \langle \sum_i \nabla_i^4 \rangle,
\end{align}
\begin{align}
\label{d1}
\langle D_1 \rangle &= \frac{\pi}{2}\alpha^2\sum_a Z_a \langle \sum_i 
\delta(\textbf{r}_{ia})\rangle,
\end{align}
where $i$ and $a$ denote electrons and nuclei, respectively. The notation $\langle 
\hat{\mathcal{O}}\rangle$ stands for the expectation value of an operator $\hat{\mathcal{O}}$ on the 
nonrelativistic clamped-nuclei wavefunction. For brevity, the consecutive terms in the above 
equation are called the mass-velocity $\langle P_4 \rangle$ and the one-electron Darwin $\langle D_1 
\rangle$ corrections. Some authors \cite{cowan76} use the name ``Cowan-Griffin correction'' for the sum of $\langle P_4
\rangle$ and $\langle D_1 \rangle$. 

Note that in the above formulation we neglected the two-electron relativistic corrections (Breit 
and two-electron Darwin contributions). For light systems they are usually at least by an order of 
magnitude smaller\cite{lesiuk19} than the one-electron corrections given by Eq. (\ref{p4}) and 
(\ref{d1}). As demonstrated further in the text, the one-electron relativistic effects contribute 
only a fraction of cm$^{-1}$ to the total interaction energy of Li$_2$. Therefore, we estimate that 
the two-electron effects are of the order of a few hundreds of cm$^{-1}$, and thus entirely 
negligible in comparison with other sources of error. An additional approximation adopted in this 
work is the neglect of spin-spin and spin-orbit interactions. The former term is very
small ($\approx0.01$ cm$^{-1}$ for all points of the potential energy curve)
as confirmed by the recent work of Minaev \cite{minaev05}, and vanishes quickly with the 
internuclear distance. The spin-orbit interaction is identically zero in the first-order 
perturbation theory since we are considering the molecular $\Sigma$ state.

The one-electron relativistic corrections were calculated analytically on the top of 
the CCSD(T) wavefunction. Contractions with the appropriate density matrices 
were accomplished by using a code written especially for this task. Because the CCSD(T) method 
performs very well for the interaction energies, we neglect the higher-order mixed 
relativistic/correlation contributions and apply no post-CCSD(T) corrections. Exemplary results of 
the calculations are given in Table \ref{trel}, where, for 
consistency, we consider the same two interatomic distances as in the preceding section. To speed 
up the calculations, we evaluated the one-electron relativistic corrections in the basis sets up to 
$l=5$ only.

From Table \ref{trel} one can see that the mass-velocity and one-electron Darwin corrections 
converge very quickly with respect to the basis set size. The results in the two largest 
basis sets are barely distinguishable. Therefore, it is not necessary to extrapolate the 
values of $\langle P_4 \rangle$ and $\langle D_1 \rangle$. The final result is simply the value 
obtained with the largest basis set and the error is estimated to be less than 5\% of the absolute 
value.

\subsection{Other corrections}
\label{subsec:qedpot}

Since the goal of the present paper is to reach the spectroscopic accuracy we have to include some further 
corrections to the potential energy curve originating from the QED and adiabatic effects. Starting with the former, the
most convenient framework to describe the QED effects in light systems is the so-called non-relativistic QED (NRQED)
theory \cite{caswell86,pachucki05}. In the NRQED the energy of the system is expanded in powers of the fine-structure
constant. The quadratic terms correspond to the aforementioned Breit-Pauli Hamiltonian and the $\alpha^3$ and $\alpha^3
\ln \alpha$ corrections are the leading-order (pure) QED effects, $E^{(3)}$. Explicit expressions for the latter are
known \cite{araki57,sucher58}, but their computation for many-electron systems is still a considerable challenge. In the
present work we adopt the following approximation to the $\alpha^3$ and $\alpha^3 \ln \alpha$ corrections
\begin{align}
\label{qed_app}
\begin{split}
E^{(3)} &\approx \frac{8\alpha}{3\pi}\left(\frac{19}{30}-2\ln \alpha - \ln k_0^{\mbox{\scriptsize Li}}\right)\langle D_1
\rangle,
\end{split}
\end{align}
where $\ln k_0$ is the Bethe logarithm \cite{bethe75,schwartz61} and $\langle D_1\rangle$ is the same as in Eq.
(\ref{d1}). This is essentially the dominant one-electron component of the complete $\alpha^3$ QED correction
(the one-electron Lamb shift). For the Bethe logarithm we adopt the atomic value, $\ln 
k_0^{\mbox{\scriptsize Li}}=5.178\,17(3)$ \cite{pachucki03}. This is a reasonable approximation 
because this quantity is usually weakly dependent on the molecular geometry
\cite{bukowski92,piszczatowski09}. For reasons similar as in the case of the relativistic 
corrections, in Eq. (\ref{qed_app}) we neglected two-electron contributions, i.e. two-electron Lamb 
shift and the Araki-Sucher correction. We assume that the approximations introduced in 
(\ref{qed_app}) result in a relative error smaller than 50\%.

Finally, let us consider the finite nuclear mass effects. The leading-order correction to the PEC due to the
nuclear motion is the so-called diagonal Born-Oppenheimer correction (or the adiabatic correction for short). It is
given by the formula \cite{komasa99,handy86}
\begin{align}
 E_{\mbox{\tiny DBOC}} = \half \sum_a \frac{1}{M_a} \langle \nabla_a \Psi_0 | \nabla_a \Psi_0 
\rangle,
\end{align}
where $a$ runs over all nuclei of the system and $M_a$ denote the nuclear masses. Unfortunately, 
calculation of the
DBOC with the basis set of STOs is not developed yet and we must resort to the GTOs in the present paper. We
have used the all-electron CCSD method to calculate the adiabatic correction \cite{gauss06} with the augmented
quadruple-zeta basis set developed by Prascher et al. \cite{prascher11} The post-CCSD corrections and basis set
incompleteness errors are neglected in this case. We assume that this introduces an error of at most 
25\%.

\subsection{Computational details}
\label{subsec:compdet}

For the record, in this section we would like to provide some additional technical details 
concerning the electronic structure calculations described above. The basis set optimizations were 
carried out by using a program written especially for this purpose. It is interfaced with the 
\textsc{Gamess} package \cite{gamess1,gamess2} which carries out the necessary CISD calculations. To 
optimize the 
well-tempering 
parameters we employed the pseudo Newton-Rhapson method with the BFGS update of the approximate 
Hessian matrix \cite{fletcher81} and numerically evaluated gradient (two-point finite difference). 
The optimizations 
were stopped when the energy difference between two consecutive cycles fell below 10 nH.

All subsequent electronic structure calculations were carried out with help of the \textsc{AcesII} 
program package \cite{aces2}. The only exception is the FCI method where the \textsc{Gamess} package was used 
and calculation of the adiabatic correction where we employed the \textsc{CFour} program \cite{cfour}. In all coupled
cluster computations we employed the restricted open-shell (RO) reference 
wavefunction. Inclusion of the relativistic corrections requires
expectation values of several operators specified in the preceding sections. Matrix elements of 
these operators were calculated directly in the STOs basis sets.
Coupled cluster density matrices were extracted from the \textsc{AcesII} package by proper manipulation of the CC 
gradients code logic.

To evaluate the complete potential energy curve we repeated the procedures described in the 
preceding sections on a grid of internuclear distances. For the nonrelativistic calculations we used the following grid:
from $R=5.5$ to $R=9.0$ the step is $R=0.25$; from $R=9.0$ to $R=14.0$ it is $R=0.5$; from $R=14.0$ to $R=25.0$ it is
$R=1.0$, and finally above $R=25.0$ the step is $R=2.5$ up to $R=40.0$ (all values are given in multiples of the
Bohr radius). Additionally, we evaluated a single point at $R=7.882$ which is close to the actual minimum of 
the potential energy curve. 
This gives a grand total of 43 points spaced from $R=5.5$ to $R=40.0$. For the relativistic corrections the grid was
slightly smaller ending at $R=30.0$. This mostly due to large cancellations occurring at large $R$ making the
calculated values less reliable.

\section{Analytic fits of the potentials}
\label{sec:fits}

\begin{table}[t]
\caption{Optimized parameters of the fit (\ref{genfit}) for the Born-Oppenheimer potential 
[$V^{\mbox{\scriptsize
BO}}(R)$] and for the adiabatic correction [$V^{\mbox{\scriptsize ad}}(R)$] (without dividing by the mass term). All
values are given in the atomic units. The symbol $X[\pm n]$ stands for $X\cdot 10^{\pm n}$. Not all digits reported are
significant.}
\begin{ruledtabular}
\label{parnrel}
\begin{tabular}{clcc}
parameter & \multicolumn{1}{c}{$V^{\mbox{\scriptsize BO}}(R)$}
          & $V^{\mbox{\scriptsize ad}}(R)$ \\[0.4ex]
\hline\\[-2.1ex]
$\alpha_1$ & $+$1.27 983[$+$00] & $+$1.87 631[$+$00] \\[0.6ex]
$\alpha_2$ & $+$2.29 122[$-$01] & $+$3.24 019[$-$01] \\[0.6ex]
$\eta$     & $+$1.02 337[$+$00] & $+$5.84 617[$-$01] \\[0.6ex]
\hline\\[-2.1ex]
$c_{01}$   & $+$1.28 843[$+$02] & $-$8.45 797[$+$00] \\[0.6ex]
$c_{11}$   & $-$9.02 013[$+$01] & $+$4.52 239[$+$00] \\[0.6ex]
$c_{21}$   & $+$2.67 910[$+$01] & $-$8.14 315[$-$01] \\[0.6ex]
$c_{31}$   & $-$3.42 393[$+$00] & $+$5.01 342[$-$02] \\[0.6ex]
$c_{41}$   & $+$2.07 665[$-$01] & ---$^{\mbox{\scriptsize b}}$ \\[0.6ex]
$c_{02}$   & $+$2.11 421[$-$03] & $+$1.95 248[$-$06] \\[0.6ex]
$c_{12}$   & $-$2.40 579[$-$04] & $-$5.40 041[$-$08] \\[0.6ex]
$c_{22}$   & $+$1.05 528[$-$05] & $-$1.43 211[$-$08] \\[0.6ex]
$c_{32}$   & $-$2.07 608[$-$07] & $+$4.64 261[$-$10] \\[0.6ex]
$c_{42}$   & $+$1.54 659[$-$09] & ---$^{\mbox{\scriptsize b}}$ \\[0.6ex]
\hline\\[-2.1ex]
$C_6$    & $+$1.39 339[$+$03]$^{\mbox{\scriptsize a}}$ & $+$1.47 084[$+$00] \\[0.6ex]
$C_8$    & $+$8.34 258[$+$04]$^{\mbox{\scriptsize a}}$ & $-$1.18 756[$+$03] \\[0.6ex]
$C_{10}$ & $+$7.37 210[$+$06]$^{\mbox{\scriptsize a}}$ & $+$4.05 449[$+$05] \\[0.6ex]
$C_{12}$ & $+$9.03 000[$+$08]$^{\mbox{\scriptsize a}}$ & ---$^{\mbox{\scriptsize b}}$ \\[0.6ex]
$C_{14}$ & $+$1.48 000[$+$11]$^{\mbox{\scriptsize a}}$ & ---$^{\mbox{\scriptsize b}}$ \\[0.6ex]
$C_{16}$ & $+$3.09 000[$+$13]$^{\mbox{\scriptsize a}}$ & ---$^{\mbox{\scriptsize b}}$ \\[0.6ex]
\end{tabular}
\begin{flushleft}\vspace{-0.2cm}
$^{\mbox{\scriptsize a}}${\small taken from Refs. \cite{yan96} and \cite{patil97}}\;
$^{\mbox{\scriptsize b}}${\small not included in the fit}\;
\end{flushleft}
\end{ruledtabular}
\end{table}

\subsection{General method}
\label{subsec:fitgen}

In order to generate results directly comparable with the experimental values, the raw \emph{ab initio} data points must
be fitted with a suitable functional form to give a smooth function of the internuclear distance, $R$. For all
contributions to the interaction energy described in the previous sections we adopt the following generic formula
\begin{align}
\label{genfit}
 V(R) = \sum_{k=1}^{N_e} e^{-\alpha_k R} \sum_{n=0}^{N_p} c_{nk} R^n - \sum_{n=3}^{N_a} \frac{C_{2n}}{R^{2n}}
f_{2n}(\eta R),
\end{align}
where $N_e$, $N_p$ and $N_a$ control the expansion length, $\alpha_k$ and $\eta$ are (nonlinear)
parameters of the fit, $c_{nk}$ are linear parameters, and $f_{2n}(\eta R)$ is the Tang-Toennies damping function
\cite{tang84}
\begin{align}
 f_{2n}(x) = 1 - e^{-x} \sum_{k=0}^{2n} \frac{x^k}{k!}.
\end{align}
The asymptotic coefficients, $C_{2n}$, in Eq. (\ref{genfit}) are either taken from more accurate theoretical
calculations or fitted (discussed further). Note that we found it unnecessary to include the repulsive Coulomb wall
(the unified atoms limit, $Z^2/R$) in the potential formula (\ref{genfit}) .

The nonlinear and linear parameters in Eq. (\ref{genfit}) are chosen to minimize weighted error of 
the fit. At each
point of the grid we are given the values of the potential, $V_k^{\mbox{\scriptsize comp}}$, and the corresponding
errors, $\delta V_k^{\mbox{\scriptsize comp}}$. The target function $\Delta$ for the optimization is 
chosen
as
\begin{align}
 \Delta^2 = \frac{1}{N_g}  \sum_{k=1}^{N_g} \left[ \frac{V_k^{\mbox{\scriptsize comp}} - V(R_k)}{\delta
V_k^{\mbox{\scriptsize comp}}} \right]^2,
\end{align}
where $V(R_k)$ is the value of the fitting function evaluated at a given grid point. We optimize the 
nonlinear
parameters by using the Powell procedure \cite{powell64}. The optimization is
stopped when
the target function varies by less than $10^{-5}$ cm$^{-1}$ between several consecutive iterations. The raw \emph{ab
initio} data ($V_k^{\mbox{\scriptsize comp}}$, $\delta V_k^{\mbox{\scriptsize comp}}$) for all components of the
 PEC are given in the Supplemental Material \cite{supplement}. A simple \textsc{Mathematica} program
\cite{math11} implementing all the fits discussed here can be
obtained from the authors upon request. 

\begin{table}[t]
\caption{Optimized parameters of the fit (\ref{genfit}) for the one-electron relativistic 
corrections -
mass-velocity [$V^{\mbox{\scriptsize P4}}(R)$] and one-electron Darwin [$V^{\mbox{\scriptsize D1}}(R)$], see Eqs.
(\ref{p4}) and (\ref{d1}), respectively, for the definitions. All values are given in the atomic units. The symbol
$X[\pm n]$ stands for $X\cdot 10^{\pm n}$.}
\begin{ruledtabular}
\label{parrel}
\begin{tabular}{cccc}
parameter & $V^{\mbox{\scriptsize P4}}(R)$
          & $V^{\mbox{\scriptsize D1}}(R)$ \\[0.4ex]
\hline\\[-2.1ex]
$\alpha_1$ & $+$1.3284[$+$00] & $+$1.3624[$+$00] \\[0.6ex]
$\alpha_2$ & $+$4.9227[$-$01] & $+$5.2275[$-$01] \\[0.6ex]
$\eta$     & $+$3.2767[$-$01] & $+$3.7855[$-$01] \\[0.6ex]
\hline\\[-2.1ex]
$c_{01}$   & $-$1.2702[$-$01] & $+$1.4557[$-$01] \\[0.6ex]
$c_{11}$   & $+$8.8288[$-$02] & $-$9.7541[$-$02] \\[0.6ex]
$c_{21}$   & $-$2.1025[$-$02] & $+$2.2358[$-$02] \\[0.6ex]
$c_{31}$   & $+$1.9023[$-$03] & $-$1.9527[$-$03] \\[0.6ex]
$c_{02}$   & $-$1.4425[$-$03] & $+$1.3737[$-$03] \\[0.6ex]
$c_{12}$   & $+$2.6974[$-$04] & $-$2.5064[$-$04] \\[0.6ex]
$c_{22}$   & $-$1.4610[$-$05] & $+$1.3126[$-$05] \\[0.6ex]
$c_{32}$   & $+$2.8967[$-$07] & $-$2.5278[$-$07] \\[0.6ex]
\hline\\[-2.1ex]
$C_6$      & $-$2.2228[$+$00] & $+$1.5773[$+$00] \\[0.6ex]
$C_8$      & $-$8.9706[$+$01] & $+$7.3228[$+$01] \\[0.6ex]
$C_{10}$   & $-$1.9637[$+$04] & $+$1.0532[$+$05] \\[0.6ex]
\end{tabular}
\end{ruledtabular}
\end{table}

\subsection{Nonrelativistic potentials}
\label{subsec:fitnrel}

An important issue in the generation of the analytic potentials is to assure that the long-range tail of PEC is
correct. Therefore, we prefer to use the asymptotic constants calculated with more accurate theoretical methods
(whenever available) rather to
rely solely on fitting to match the data points. Fortunately, reliable values of the first three nonrelativistic
asymptotic constants ($C_6$, $C_8$, $C_{10}$) were given by Yan et al. \cite{yan96} These values were obtained
from
variational wave functions in Hylleraas basis sets and are all accurate to better than one part per thousand. For
the higher
asymptotic constants ($C_{2n}$ with $n>5$) the data in the literature are not as abundant. Remarkably, Patil
et al. \cite{patil97} report values of the asymptotic constants up to $n=12$. Their values are progressively less
reliable with increasing $n$. For example, we find that the error in $C_6$ is only about 0.3\% compared with the work
of Yan et al. \cite{yan96} but rises to almost 2\% for $C_{10}$. Therefore, we adopt the values of
$C_{12}$, $C_{14}$
and $C_{16}$ from Ref. \cite{patil97} and neglect the higher-order inverse powers of $R$ in Eq. (\ref{genfit}). We
checked that the
inclusion of terms beyond $C_{16}$ changes the results only marginally. The same is true for the asymptotic
terms such as $C_{11}/R^{11}$ (resulting from higher-order perturbation theory) which can be safely neglected at this
point.

Concerning the adiabatic correction, the corresponding asymptotic constants are not available for lithium. Despite
explicit expressions for these coefficients are available in the literature \cite{mitek12}, their calculation is
complicated and has been achieved only for one- and two-electron systems thus far. Therefore, we have no other option
but to obtain the asymptotic constants $C_{2n}^{\mbox{\scriptsize ad}}$ by fitting. We find that inclusion of the
first three coefficients is sufficient to provide a reasonable accuracy.

Overall, the fitting function (\ref{genfit}) with $N_e=2$, $N_p=3\;\mbox{or}\;4$, and $N_a\leq8$ provides a satisfactory
representation of the raw \emph{ab initio} data, both for the Born-Oppenheimer results [$V^{\mbox{\scriptsize BO}}(R)$,
$N_p=4$, $N_a=8$] and for the adiabatic correction [$V^{\mbox{\scriptsize ad}}(R)$, $N_p=3$, $N_a=5$]. Both fits
contain 10 linear and 3 nonlinear parameters which is a modest amount compared to about 40 points of the raw
\emph{ab initio} data. The fitting errors are by an order of magnitude smaller than the estimated
uncertainty of the corresponding theoretical calculations. Only one or two points are exceptional in this
respect, but the error is still well within the acceptable range. Optimized  parameters of the 
Born-Oppenheimer and
adiabatic potentials are given in Table \ref{parnrel}. Note that the adiabatic correction fitting error is larger than
for the BO potential [cf. Table \ref{fiterr}] but this mostly due to increased relative errors $\delta
V_k^{\mbox{\scriptsize comp}}$ and smaller number of fitting parameters.

\begin{table}[t]
\caption{Root mean square deviations (in cm$^{-1}$) and maximum absolute deviations (percentage-wise) of the
fitted values from the raw data points. The symbol $X[\pm n]$ stands for $X\cdot 10^{\pm n}$.}
\begin{ruledtabular}
\label{fiterr}
\begin{tabular}{lcc}
 & rms error & max error (\%) \\[0.6ex]
\hline\\[-1.8ex]
 $V^{\mbox{\scriptsize BO}}(R)$ & 1.8[$-$01] & 3.0[$-$01] \\[0.6ex]
 $V^{\mbox{\scriptsize D1}}(R)$ & 3.6[$-$05] & 6.7[$-$02] \\[0.6ex]
 $V^{\mbox{\scriptsize P4}}(R)$ & 8.1[$-$05] & 6.1[$-$02] \\[0.6ex]
 $V^{\mbox{\scriptsize ad}}(R)$ & 6.0[$-$03] & 4.7[$+$00] \\
\end{tabular}
\end{ruledtabular}
\end{table}

\subsection{Relativistic effects}
\label{subsec:fitrel}

Analytic potentials corresponding to the one-electron relativistic effects were obtained in a similar fashion as
for the
adiabatic correction. The mass-velocity [Eq. (\ref{p4})] and one-electron Darwin [Eq. (\ref{d1})] terms were separately
represented in the form given by Eq. (\ref{genfit}) with $N_e=2$, $N_p=3$, $N_a=5$. The optimized 
parameters are given
in Table \ref{parrel}. For convenience, in both cases we have included the factor of $\alpha^2$ into the coefficients.

Note that the last asymptotic constant ($C_{10}$) in both fits optimized to a surprisingly large 
value. We
believe that this result should be treated cautiously. Whilst the first two asymptotic coefficients are
reasonably stable with respect to various modifications of the fitting formula, the last one depends significantly on
the adopted parametrization. In order to stabilize this quantity one would need to include more 
asymptotic terms, but
because of the risk of over-parametrization, we decided not to do it. Therefore, the obtained values 
of $C_{10}$ should
not be used as a reference for other methods. The same conclusion is probably valid for the fit of the adiabatic
correction described in the previous section.

The accuracies of the fitting functions for are summarized in Table \ref{fiterr}. More detailed data 
are given in
Supplemental Material \cite{supplement}. This includes explicit listing of the raw \emph{ab initio} values at each
point and the corresponding errors.

\section{Spectroscopic data}
\label{sec:spectro}

\begin{figure}[b]
\includegraphics[scale=1.0]{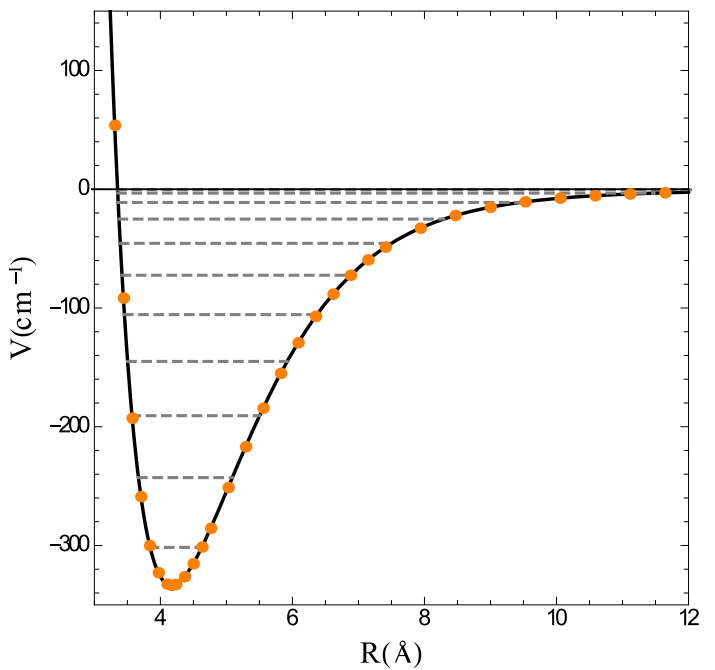}
\caption{\label{fig:curveli} Complete potential energy curve for the for the $a^3\Sigma_u^+$ state 
of $^{7,7}$Li$_2$ (solid black line);
orange dots are the actual \emph{ab initio} data points. The horizontal dashed lines are energies of the $J=0$
vibrational levels. The horizontal black solid line denotes the onset of continuum.}
\end{figure}

In order to generate the spectroscopic data we add up all components of the PEC described above (BO, adiabatic,
relativistic and QED). The final PEC is illustrated in Fig. \ref{fig:curveli}. Based on the complete 
curve we calculate
the relevant molecular parameters. The total binding
energy (i.e. the well depth, $D_e$) and the equilibrium internuclear distance ($R_e$) are obtained by finding the
minimum of the fitted PEC. The harmonic vibrational frequency is defined as
\begin{align}
 \omega_e^2 = \frac{1}{\mu} \left.\left( \frac{\partial^2 V}{\partial R^2} \right)\right|_{R_e},
\end{align}
in the atomic units, where $\mu$ is the reduced mass of an isotopomer. We consider two stable isotopes
of lithium ($^6$Li and $^7$Li) with the atomic masses equal to
\begin{align}
 m(^6\mbox{Li}) = \mbox{6.015 123 u},\\
 m(^7\mbox{Li}) = \mbox{7.016 005 u},
\end{align}
according to the recent compilation \cite{audi03}.

In order to find the rovibrational wavefunctions ($\Psi_{\nu J}$) and energies ($E_{\nu J}$) we solve the nuclear
(radial) Schr\"{o}dinger equation within the adiabatic approximation
\begin{align}
\label{radschr}
\begin{split}
 \left[-\frac{1}{2\mu}\frac{d^2}{dR^2} + D_e + V(R) + \frac{J(J+1)}{2\mu R^2} - E_{\nu J}\right] \Psi_{\nu J}(R) = 0,
\end{split}
\end{align}
where $J$ is the rotational quantum number. Note that we have added the well-depth ($D_e$) to the 
left-hand-side
of Eq. (\ref{radschr}). This makes all $E_{\nu J}$ positive by convention and their values grow with the increasing
values of $\nu$ and $J$. Further in the paper we are mostly concerned with the lowest rotational state ($J=0$) and thus
adopt the notation $E_{\nu} := E_{\nu 0}$. Finally, the dissociation energy is defined as a sum of
the interaction energy and the zero-point vibrational energy, $D_0 = D_e + E_{\nu=0}$.

\begin{table}[t]
\caption{Molecular parameters of the $a^3\Sigma_u^+$ state of $^{6,6}$Li$_2$ and $^{7,7}$Li$_2$. See the main text for
precise definitions of the listed quantities. All values are given in cm$^{-1}$, apart 
from $R_e$ which are given in \AA{}ngstr\"{o}ms, \AA{}.}
\begin{ruledtabular}
\label{molpar}
\begin{tabular}{l|cccc}
 & $D_e$ & $R_e$ & $D_0$ & $\omega_e$ \\[0.6ex]
\hline\\[-1.8ex]
 &  \multicolumn{4}{c}{$^{6,6}$Li$_2$}  \\[0.6ex]
\hline\\[-1.8ex]
 this work\;\;\;       & 333.68(30) & 4.1688 & 299.13 & 71.05\;\, \\[0.6ex]
 Ref. \cite{dattani11} & 333.778(8) & 4.170038(30) & --- & 70.65$^{\mbox{\scriptsize a}}$ \\[0.6ex]
\hline\\[-1.8ex]
 &  \multicolumn{4}{c}{$^{7,7}$Li$_2$}  \\[0.6ex]
\hline\\[-1.8ex]
 this work\;\;\;       & 333.69(30) & 4.1687 & 301.61 & 65.78\;\, \\[0.6ex]
 Ref. \cite{dattani11} & 333.758(7) & 4.17005(3) & --- & 65.42$^{\mbox{\scriptsize a}}$ \\[0.6ex]
 Ref. \cite{linton99}  & 333.69(10) & 4.173 & 301.829(15) & --- \\[0.6ex]
\end{tabular}
 \begin{flushleft}\vspace{-0.2cm}
 $^{\mbox{\scriptsize a}}${\small not reported originally in Ref. \cite{dattani11}; extracted by taking the second
derivative of the final potential}
 \end{flushleft}
\end{ruledtabular}
\end{table}

In Table \ref{molpar} we report the calculated \emph{ab initio} values of the molecular parameters ($D_e$, $R_e$, $D_0$,
$\omega_e$) for both isotopomers of the lithium dimer. The error of $D_e$ was estimated by interpolating the theoretical
errors at several neighboring grid points. Let us compare our results with the most recent 
experimental values of
Linton et
al. \cite{linton99} and with a very reliable 17-parameter Morse/long-range potential of Dattani and Le Roy
\cite{dattani11}. The agreement with these
values is remarkably good. For example, our $D_e$ for the isotopomer $^{7,7}$Li$_2$ differs from the results of Refs.
\cite{linton99} and \cite{dattani11} by only 0.01 and 0.07 cm$^{-1}$, respectively, while our estimated error is about
0.3 cm$^{-1}$ at the bottom of the well. The same conclusion is valid for the
dissociation energy, $D_0$. This suggest that our error estimations are indeed quite conservative, at least in the
regions close to the minimum of the potential. A similarly good agreement is found for the remaining molecular
parameters.

\begin{table}[t]
\caption{Vibrational energy levels ($J=0$) for the $a^3\Sigma_u^+$ state of $^{7,7}$Li$_2$. The vibrational energies
($E_{\nu}$) are given in cm$^{-1}$, and the classical turning points ($R_{\mbox{\scriptsize 
min}}^\nu$, $R_{\mbox{\scriptsize max}}^\nu$) in \AA{}ngstr\"{o}ms, \AA{}. The minimum of PEC 
corresponds to the zero energy. The last two rows are the maximum and root-mean-square errors with 
respect to the experimental data
\cite{linton99}.}
\begin{ruledtabular}
\label{viben}
\begin{tabular}{c|rrrrrr}
 & \multicolumn{3}{c}{this work} & \multicolumn{3}{c}{Ref. \cite{linton99}} \\[0.6ex]
\hline\\[-2.2ex]
 $\nu$ & $E_{\nu}$\;\; & $R_{\mbox{\scriptsize min}}^\nu$ & $R_{\mbox{\scriptsize max}}^\nu$ & $E_{\nu}$\;\;\; &
$R_{\mbox{\scriptsize min}}^\nu$ & $R_{\mbox{\scriptsize max}}^\nu$ \\[0.6ex]
\hline\\[-2.2ex]
 0 & 32.06  & 3.844 & 4.627 & 31.857 & 3.846 & 4.630 \\[0.6ex]
 1 & 90.83  & 3.668 & 5.090 & 90.453  & 3.668 & 5.092 \\[0.6ex]
 2 & 142.94 & 3.570 & 5.502 & 142.523 & 3.571 & 5.503 \\[0.6ex]
 3 & 188.65 & 3.504 & 5.920 & 188.240 & 3.505 & 5.922 \\[0.6ex]
 4 & 228.07 & 3.455 & 6.371 & 227.679 & 3.458 & 6.373 \\[0.6ex]
 5 & 261.24 & 3.419 & 6.882 & 260.837 & 3.422 & 6.885 \\[0.6ex]
 6 & 288.11 & 3.392 & 7.496 & 287.665 & 3.395 & 7.501 \\[0.6ex]
 7 & 308.55 & 3.373 & 8.293 & 308.098 & 3.377 & 8.297 \\[0.6ex]
 8 & 322.55 & 3.361 & 9.453 & 322.155 & 3.365 & 9.441 \\[0.6ex]
 9 & 330.39 & 3.354 & 11.476 & 330.170 & 3.358 & 11.392 \\[0.6ex]
10 & 333.32 & 3.352 & 16.478 & 333.269 & 3.356 & 16.052 \\[0.6ex]
\hline\\[-2.2ex]
$\delta_{\mbox{\scriptsize max}}$ & 0.45 & 0.004 & 0.424 & \multicolumn{1}{c}{---} & \multicolumn{1}{c}{---} &
\multicolumn{1}{c}{---}
\\[0.6ex]
$\delta_{\mbox{\scriptsize rms}}$ & 0.34 & 0.003 & 0.130 & \multicolumn{1}{c}{---} & \multicolumn{1}{c}{---} &
\multicolumn{1}{c}{---}
\\[0.6ex]
\end{tabular}
\end{ruledtabular}
\end{table}

The radial nuclear Schr\"{o}dinger equation (\ref{radschr}) was solved with help of the discrete variable
representation (DVR) method \cite{colbert92}. The obtained vibrational energy levels ($J=0$) are listed in Table
\ref{viben} and compared
with the experimental values of Linton et al. \cite{linton99}. Additionally, we calculate the classical turning points
($R^\nu$)
defined as solutions of the following implicit equations
\begin{align}
\label{turn}
 D_e + V(R^\nu) = E_{\nu}.
\end{align}
For each $\nu$ we have two solutions of Eq. (\ref{turn}), denoted
$R_{\mbox{\scriptsize min}}^\nu$ and $R_{\mbox{\scriptsize max}}^\nu$, and both of them are listed in Table
\ref{viben}. 

One can see an excellent agreement between the theoretical and experimental vibrational energy levels, Table
\ref{viben}. The maximum absolute deviation is found for $\nu=7$ and amounts to about 0.4 cm$^{-1}$. On average,
the deviation is of the order of 0.3 cm$^{-1}$. Let us point out that resolution of the spectroscopic data of Linton
et al. is about 0.1 cm$^{-1}$, so that the actual error of our calculations can be even smaller. Moreover, our \emph{ab
initio} values are more accurate than reported recently by Lau et al. \cite{lau16}
based on a semi-empirical model potential. Their data exhibits the maximum $E_{\nu}$ deviation of about 1.5
cm$^{-1}$ if they use the accurate $\omega_e$ in the potential. By relaxing the value of $\omega_e$ by about
1\% the accuracy improves to about 0.5 cm$^{-1}$ on the average, but this may be due to a fruitful cancellation of
errors. In fact, our results support the semiempirical value of $\omega_e$. Let us also point out that our potential
reproduces the binding energy of the last vibrational level with surprising accuracy. While the experimentally derived
value is 12.47$\,\pm\,$0.04 GHz \cite{abraham95} the PEC developed in this work gives 10.5 GHz.

Let us now turn our attention to the theoretical description of the Li$-$Li scattering process. The 
main goal is to evaluate
the $s$-wave scattering length ($a$) for two lithium atoms in the ground state from the first-principles PEC developed
in this work. This can be accomplished by solving the radial Schr\"{o}dinger equation (\ref{radschr}) with $J=0$ at
zero energy \cite{moszynski03}. It is well known that for large $R$ the solutions $\Psi_{\mbox{\scriptsize E}=0}(R)$
behave asymptotically as a linear function \cite{mott65,landau65}
\begin{align}
\label{scatasym}
 \Psi_{\mbox{\scriptsize E}=0}(R) \rightarrow C\left(R-a\right)+\ldots,
\end{align}
where $a$ is the desired scattering length. Very sophisticated methods for numerical calculation of $a$ were
presented \cite{gutierrez84,marinescu94,szmytkowski95,meshkov11,hon03}, but our case is not 
particularly technically
challenging and we adopt the following simplistic procedure. First, we propagate the radial Schr\"{o}dinger equation at
zero energy up to very large $R$ ($\approx 10^5$). The initial conditions are $\Psi_{\mbox{\scriptsize E}=0}(R_0)=0$,
where $R_0$ is deep within the repulsive wall, and an arbitrary value of the derivative at $R_0$. Next, we continue the
asymptotic straight line (\ref{scatasym}) to the point where it crosses the $r$-axis. By the virtue of Eq.
(\ref{scatasym}) this point corresponds to the value of $a$.

The $s$-wave scattering length for the $^{7,7}$Li$_2$ isotopomer calculated from the PEC developed in this work is
$-9.2$ a.u. This is by a factor of three too small compared with the experimental result of Abraham et al.
\cite{abraham95} who report $-27.3\,\pm\,0.8$ a.u. Despite this deviation is large we note that the sign of the
scattering length calculated by us is correct. This is sufficient to predict the stability of the corresponding
Bose-Einstein condensate \cite{huang87,stoof94}. Moreover, the rough magnitude of the scattering length is also correct
which makes it useful for other predictions \cite{hess87,doyle91}. To predict $a$ with the accuracy of a few percents
the errors in PEC must be reduced probably by an order of magnitude. We believe that this is possible in a
foreseeable future.

\section{Conclusions}
\label{sec:concl}

In this paper we have developed a new \emph{ab initio} potential energy curve for the $a^3\Sigma_u^+$ state of lithium
dimer. To bring down the accuracy to the sub-cm$^{-1}$ regime we have employed state-of-the-art 
techniques of the
electronic structure theory. In particular, large (double to sextuple zeta) one-electron basis sets composed of
Slater-type orbitals have been developed specifically for the present purposes. The Born-Oppenheimer potential has been
calculated by using a composite scheme utilizing high-order coupled cluster and full CI methods. 
Moreover, we have
included several minor corrections to account for the the adiabatic, relativistic, and QED effects.

The computed \emph{ab initio} data points have been fitted with theoretically motivated analytic functions.
When available, we employed van der Waals asymptotic constants $C_n$ obtained from the most accurate theoretical
methods. By solving the nuclear Schr\"{o}dinger equation we have obtained the molecular parameters ($D_e$, $D_0$,
$\omega_e$ etc.) for this system, as well as the corresponding vibrational energy levels, which are directly comparable
with the experimental data. For example, the bond dissociation energy determined by us ($D_0=301.61$ cm$^{-1}$) differs
by only about 0.2 cm$^{-1}$ from the empirical values reported by Linton et al. \cite{linton99} We have also reproduced
all eleven bound vibrational levels with an accuracy of $0.2-0.4\,$cm$^{-1}$. In particular, the 
position of the last
vibrational level has been predicted to within 2 GHz or 15\% of the experimental value. Crucially, all these results
have been obtained without prior adjustment to match the empirical values.

The data presented in this paper are probably the most accurate \emph{ab initio} results available 
for this system in
the literature thus far. Moreover, this paper constitutes a proof that Slater-type orbitals can now be used routinely
in calculations for the diatomic systems with large basis sets (up to several hundred functions) and are capable of
providing spectroscopically accurate results.

\begin{acknowledgments}
ML acknowledges the support by the Foundation for Polish Science (FNP) and by the Polish National 
Agency of Academic Exchange through the Bekker programme No. PPN/BEK/2019/1/00315/U/00001. 
RM was supported by the Polish National Science Center through Grant No. 2016/21/B/ST4/03877. 
Computations presented in this research were carried out with the support of 
the Interdisciplinary Center for Mathematical and Computational Modeling (ICM) at the University 
of Warsaw, grant number G59-29. We would like to thank Ms. Iwona Majewska for providing a DVR
program to solve the nuclear Schr\"{o}dinger equation.
\end{acknowledgments}

\end{document}